\documentclass[twocolumn,superscriptaddress,floats,prd,nofootinbib,showpacs]{revtex4}

\usepackage{amsmath,amssymb,amsfonts}
\usepackage{graphicx}

\def\be{\begin{equation}}
\def\ee{\end{equation}}
\def\ba{\begin{eqnarray}}
\def\ea{\end{eqnarray}}
\def\nn{\nonumber}

\newcommand{\Pl}{\ell_\textrm{Pl}} 
\newcommand{\mubar}{{\bar \mu}} 
\newcommand{\abs}[1]{{\left|{#1}\right|}} 

\newcommand{\eqnref}[1]{(\ref{#1})}
\newcommand{\figref}[1]{Fig.~\ref{#1}}

\begin{document}


\title{Cosmological inflation driven by holonomy corrections of loop quantum cosmology}
\author{Dah-Wei Chiou}
\email{chiou@gravity.psu.edu}
\affiliation{
Department of Physics, Beijing Normal University, Beijing 100875, China}
\author{Kai Liu}
\email{liukai@mail.bnu.edu.cn}
\affiliation{
Department of Physics, Beijing Normal University, Beijing 100875, China}

\begin{abstract}
At the level of heuristic effective dynamics, we investigate the cosmological inflation with holonomy corrections of loop quantum cosmology (LQC) in the $k=0$ Friedmann-Robertson-Walker model with a single inflaton field subject to a simple potential. In the symmetric bouncing scenario of LQC, the condition for occurrence of the quantum bounce naturally and uniquely fixes the initial conditions at the bouncing epoch. Around the quantum bounce, the universe undergoes a short super-inflationary phase, which drives the inflaton field to its potential hill and thus sets the proper initial conditions for the standard slow-roll inflation. Between the super-inflation and the standard inflation, there is a non-inflationary phase, which violates the slow-roll condition. The violation of slow roll is expected to give some suppression on the low angular power spectrum of the cosmic microwave background and different orders of holonomy corrections shall yield different suppressions.
\end{abstract}

\pacs{98.80.Qc, 03.65.Sq, 04.60.Pp}

\maketitle


\section{Introduction}
The status of loop quantum cosmology (LQC) has undergone lively progress and has become an active area of research over the previous years. Particularly, the comprehensive formulation of LQC in the spatially flat and isotropic ($k=0$) Friedmann-Robertson-Walker (FRW) model has been constructed in detail, giving a solid foundation for the quantum theory and revealing that the big bang singularity is resolved and replaced by the \emph{quantum bounce}, which bridges the present universe with a preexisting one \cite{Ashtekar:2006rx,Ashtekar:2006uz,Ashtekar:2006wn}. Similar results are also affirmed for a variety of extended models \cite{Ashtekar:2006es,Vandersloot:2006ws,Chiou:2006qq,Chiou:2007mg,Chiou:2007sp}.

Meanwhile, it has been shown that loop quantum corrections on the eigenvalues of the density operator drive super-inflationary expansion \cite{Bojowald:2002nz,Bojowald:2003mc,Date:2004yz}. Although the short super-inflation alone cannot yield sufficient $e$-foldings, in the presence of the inflaton potential $V(\phi)$, the quantum corrections can drive the inflaton field $\phi$ up to its potential hill, thus setting the initial conditions for the standard slow-roll inflation and in principle leaving an indirect signature on the large scales in the cosmic microwave background (CMB) \cite{Tsujikawa:2003vr}. In addition to corrections on the density, LQC possesses the other kind of quantum effects --- holonomy corrections. Super-inflation due to holonomy corrections and the effects on inflationary models have also been studied \cite{Artymowski:2008sc,Mielczarek:2009zw,Grain:2009kw}.

However, despite being an attractive idea that the super-inflation sets the initial conditions for the standard inflation, all of the scenarios studied so far still involve some tuning parameters (e.g. $\dot{\phi}$) to start with in the super-inflation or pre-bounce era. In this paper, we consider the quantum bounce due to holonomy corrections\footnote{It has been shown that holonomy corrections, rather than corrections on the density, are responsible for the quantum bounce and in fact it is a good approximation to neglect the corrections on the density \cite{Ashtekar:2007em}.} and study the inflation initiated by the bounce in a \emph{symmetric} bouncing scenario (i.e. the solution in which the pre-bounce and post-bounce universes are exactly the mirrored solutions to each other). In the symmetric scenario, once $V(\phi)$ is given, the condition for occurrence of the quantum bounce will naturally and uniquely set the initial conditions at the bouncing epoch and thus no extra parameters are tunable except for those describing the potential. To be more generic, we also include higher orders of holonomy corrections as suggested and studied in \cite{Chiou:2009hk,Chiou:2009yx}.

At the level of heuristic effective dynamics, we investigate the cosmological evolution with holonomy corrections in the $k=0$ FRW model with a single scalar inflaton field $\phi$ subject to a potential. In the vicinity of the bounce, there is a super-inflationary phase, which drives $\phi$ to its potential hill and thus sets the proper initial conditions for the standard slow-roll inflation. Between the super-inflation and the standard inflation, there is a non-inflationary phase, which violates the slow-roll condition. The scenario with the super-inflationary, non-inflationary and standard inflationary phases in succession is in a very similar fashion to that in \cite{Tsujikawa:2003vr}, but the super-inflationary phase is now caused by holonomy corrections, instead of the corrections on the density operator. By choosing an appropriate mass of $\phi$, the super-inflation can raise $\phi$ high enough to achieve sufficient $e$-foldings.

This scenario is robust for different orders of holonomy corrections, as different values of the order $n$ only slightly modify the maximal value of $\phi$ and the number of $e$-foldings. Furthermore, the quantum bounce and thus the super-inflationary phase are shorter and more abrupt as $n$ increases, in accord with the analysis of \cite{Chiou:2009hk}.

As discussed in \cite{Tsujikawa:2003vr}, the violation of slow roll is expected to give some suppression on the low multipole spectrum of the CMB anisotropy. Different orders $n$ shall yield different suppressions and in principle the very low angular amplitudes can be strongly suppressed in the limit $n\rightarrow\infty$. We hope to investigate the details in the future.

\section{Higher order holonomy corrections}\label{sec:holonomy corrections}
In the $k=0$ FRW model with a scalar inflaton field $\phi$ subject to a potential $V(\phi)$, the classical Hamiltonian constraint is given by
\ba\label{eqn:classical H}
H_\mathrm{cl}&=&H_\textrm{grav}+H_\phi\nn\\
&=&-\frac{3}{8\pi G\gamma^2}\,c^2\sqrt{p}+\frac{p_\phi^2}{2p^{3/2}}+p^{3/2}V(\phi),
\ea
where $c$ and $p$ are the Ashtekar variables satisfying the canonical relation
\be\label{eqn:c and p}
\{c,p\}=\frac{8\pi G\gamma}{3},
\ee
$\gamma$ is the Barbero-Immirzi parameter, and $p_\phi$ is the conjugate momentum of the inflaton field $\phi(\vec{x},t)=\phi(t)$ satisfying
\be
\{\phi,p_\phi\}=1.
\ee
The Ashtekar variable $p$ is related to the scale factor $a$ by
\be
p=L^2a^2
\ee
with $L$ being the coordinate length of the finite sized cubic cell chosen to make the Hamiltonian finite (see \cite{Chiou:2007mg}).

One of the essential features of LQC is that the connection variable $c$ does not exist and should be replaced by holonomies. The simple prescription is to replace $c$ with $\sin(\mubar c)/\mubar$ by introducing the discreteness variable $\mubar$. As suggested and studied in \cite{Chiou:2009hk,Chiou:2009yx}, a more sophisticated prescription to implement the underlying structure of LQC is to replace $c$ with $n$th order holonomized connection variable defined as
\be\label{eqn:holonomized c}
c_h^{(n)}:=\frac{1}{\mubar}\sum_{k=0}^{n} \frac{(2k)!}{2^{2k}(k!)^2(2k+1)}\,
{(\sin \mubar c)}^{2k+1},
\ee
which can be made arbitrarily close to $c$ (as $n\rightarrow\infty$) but remains a function of the holonomy $\sin \mubar c$ and the discreteness variable $\mubar$. The simple prescription now reads as $c_h^{(n=0)}$.

The Hamiltonian constraint with holonomy corrections up to the $n$th order now reads as
\be\label{eqn:H mubar n}
H_\mubar^{(n)}
=-\frac{3}{8\pi G\gamma^2}\,(c_h^{(n)})^2
\sqrt{p}+\frac{p_\phi^2}{2p^{3/2}}
+p^{3/2}V(\phi).
\ee
Particularly, in the improved dynamics suggested by \cite{Ashtekar:2006wn}, the discreteness variable $\mubar$ is given by
\be\label{eqn:mubar scheme}
\mubar=\sqrt{\frac{\Delta}{p}}\,,
\ee
where $\Delta$ is the area gap in the full theory of LQG and $\Delta=2\sqrt{3}\,\pi\gamma\Pl^2$ for the standard choice (but other choices are also possible) with $\Pl:=\sqrt{G\hbar}$ being the Planck length.

Equation \eqnref{eqn:c and p} follows
\begin{subequations}
\ba
\{c,c_h^{(n)}\}
&=&\frac{8\pi G\gamma}{3\mubar}\frac{\partial\mubar}{\partial p}
\left[
\cos(\mubar c)\,\mathfrak{S}_n(\mubar c)\,c-c_h^{(n)}
\right],\quad\\
\{p,c_h^{(n)}\}
&=&-\frac{8\pi G\gamma}{3}\cos(\mubar c)\,\mathfrak{S}_n(\mubar c),
\ea
\end{subequations}
where
\begin{subequations}
\ba
\mathfrak{S}_n(\mubar c)&:=&
\sum_{k=0}^{n} \frac{(2k)!}{2^{2k}(k!)^2}\,{(\sin \mubar c)}^{2k}\\
&
\mathop{\longrightarrow}\limits_{n \rightarrow\infty}
&
{\abs{\cos \mubar c}}^{-1}.
\ea
\end{subequations}

\section{Heuristic effective dynamics}\label{sec:heuristic dynamics}
At the level of heuristic effective dynamics, the evolution is solved as if the dynamics was classical but governed by the new Hamiltonian \eqnref{eqn:H mubar n} with holonomy corrections up to the $n$th order. For the case of $n=0$ and $V(\phi)=0$, it has been shown that the heuristic treatment gives a very good approximation for the quantum evolution of LQC, and the bouncing scenario of the effective solution gives the absolute upper bound for the matter density \cite{Bojowald:2006gr,Chiou:2008bw}. For $n>0$ and $V(\phi)=0$, the quantum theory of LQC yields the quantum evolution which is very close to that obtained by heuristic effective dynamics (compare \cite{Chiou:2009yx} with \cite{Chiou:2009hk}).\footnote{\label{footnote:high order}Moreover, in the limiting case $n\rightarrow\infty$, the heuristic solution exhibits some discontinuity (kink) \cite{Chiou:2009hk}, which is well smoothed by quantum fluctuations as can be seen in \cite{Chiou:2009yx}. The trajectory of the expectation value of $p$ in the quantum theory at the order $n=\infty$ behaves like the solution obtained by heuristic effective dynamics with very huge but finite $n$.} For the generic $n$ with $V(\phi)\neq0$, the reliability remains to be justified, but we believe that the comparison between \cite{Chiou:2009hk} and \cite{Chiou:2009yx} still holds and the heuristic analysis can again provide good ideas of what the quantum evolution looks like.

By choosing the lapse function $N=p^{3/2}$ associated with the new time variable $t'$ via $dt'=N^{-1}dt$ ($t$ is the proper time), the modified Hamiltonian \eqnref{eqn:H mubar n} is rescaled and simplified as
\be
H_\mubar^{(n)'}=-\frac{3}{8\pi G\gamma^2}\,(c_h^{(n)})^2p^2+\frac{p_\phi^2}{2}
+p^3V(\phi).
\ee
The effective equations of motion are then given by the Hamilton's equations:
\begin{subequations}\label{eqn:Hamilton eqs}
\ba
\label{eqn:eom 1}
\frac{dp_\phi}{dt'}&=&\{p_\phi,H_\mubar^{(n)'}\}=-p^3V'(\phi)\\
\label{eqn:eom 2}
\frac{d\phi}{dt'}&=&\{\phi,H_\mubar^{(n)'}\}=p_\phi,\\
\label{eqn:eom 3}
\frac{dc}{dt'}&=&\{c,H_\mubar^{(n)'}\}=-\frac{2}{\gamma}(c_h^{(n)})^2p\nn\\
&&\qquad -\frac{2}{\gamma\mubar}\frac{\partial \mubar}{\partial p}
\left[\cos(\mubar c)\,\mathfrak{S}_n(\mubar c)\,c-c_h^{(n)}\right]c_h^{(n)}p^2\nn\\
&&\qquad+\,8\pi G\gamma\, p^2V(\phi),\\
\label{eqn:eom 4}
\frac{dp}{dt'}&=&\{p,H_\mubar^{(n)'}\}
=\frac{2}{\gamma}\cos(\mubar c)\,\mathfrak{S}_n(\mubar c)\,c_h^{(n)}p^2,
\ea
\end{subequations}
and the constraint that the Hamiltonian must vanish $H_\mubar^{(n)'}=0$:
\be\label{eqn:eom 5}
\frac{p_\phi^2}{2}+p^3V(\phi)=\frac{3}{8\pi G\gamma^2}\,(c_h^{(n)})^2p^2.
\ee

Equations \eqnref{eqn:eom 4} and \eqnref{eqn:eom 5} yield the modified Friedmann equation for the Hubble rate:
\ba\label{eqn:Friedmann eq}
H&:=&\frac{\dot{a}}{a}=\frac{1}{\sqrt{p}}\frac{d\sqrt{p}}{dt}
=\frac{1}{2p^{5/2}}\frac{dp}{dt'}\nn\\
&=&\frac{1}{\gamma}\cos(\mubar c)\,
\mathfrak{S}_n(\mubar c)\,c_h^{(n)}\,\frac{1}{\sqrt{p}}\nn\\
&=&\left(\frac{8\pi G}{3}\right)^{1/2}\cos(\mubar c)\,\mathfrak{S}_n(\mubar c)
\,\rho_\phi^{1/2},
\ea
where the matter density of $\phi$ is given by
\be
\rho_\phi=\frac{p_\phi^2}{2p^3}+V(\phi)=\frac{\dot{\phi}^2}{2}+V(\phi).
\ee
We can also compute
\ba
\ddot{\phi}&:=&\frac{d^2\phi}{dt^2}
=\frac{1}{p^{3/2}dt'}\left(\frac{1}{p^{3/2}}\frac{d\phi}{dt'}\right)
=\frac{1}{p^{3/2}dt'}\left(\frac{p_\phi}{p^{3/2}}\right)\nn\\
&=&-V'(\phi)-\frac{3p_\phi}{\gamma p^2}
\cos(\mubar c)\,\mathfrak{S}_n(\mubar c)\,c_h^{(n)}\nn\\
&=&-V'(\phi)-\frac{3\dot{\phi}}{\gamma\sqrt{p}}
\cos(\mubar c)\,\mathfrak{S}_n(\mubar c)\,c_h^{(n)},
\ea
which leads to the Klein-Gordon equation:
\be\label{eqn:KG eq}
\ddot{\phi}+3H\dot{\phi}+V'(\phi)=0.
\ee
The quantum corrected Raychaudhuri equation follows from \eqnref{eqn:Friedmann eq} and \eqnref{eqn:KG eq}:
\ba\label{eqn:Raychaudhuri eq}
\dot{H}&=&-4\pi G\left[\cos(\mubar c)\,\mathfrak{S}_n(\mubar c)\right]^2\dot{\phi}^2\\
&&\ +H \left[\cos(\mubar c)\,\mathfrak{S}_n(\mubar c)\right]^{-1}
\frac{d}{dt}\left[\cos(\mubar c)\,\mathfrak{S}_n(\mubar c)\right].\nn
\ea

In the classical regime, we have $\mubar c\ll 1$ and thus $\cos(\mubar c)\rightarrow 1$, $\sin(\mubar c)\rightarrow 0$, $\mathfrak{S}_n(\mubar c)\rightarrow 1$ and $c_h^{(n)}\rightarrow c$; therefore, the above equations all reduce to their classical counterparts. In the backward evolution, the quantum corrections are more and more significant as $\mubar c$ becomes appreciable. Eventually, $p$ gets bounced at the epoch when $\cos(\mubar c)$ in \eqnref{eqn:Friedmann eq} flips signs. The exact point of the quantum bounce is given by $\cos(\mubar c)=0$ (i.e. $\mubar c=\pi/2$). By \eqnref{eqn:holonomized c}, this happens when
\be
c_h^{(n)}\mubar=\sum_{k=0}^{n} \frac{(2k)!}{2^{2k}(k!)^2(2k+1)}
=:\mathfrak{F}_n
\ee
(note that $\mathfrak{F}_n\rightarrow \pi/2$ as $n\rightarrow\infty$), or equivalently, by \eqnref{eqn:eom 5}, when
\be
\label{eqn:rho crit}
\rho_\phi=\rho_\textrm{crit}^{(n)}:=3\mathfrak{F}_n^2\,\rho_\textrm{Pl}
\ee
with the Planckian density defined as
\be\label{eqn:rho Pl}
\rho_\textrm{Pl}:=(8\pi G \gamma^2\Delta)^{-1},
\ee
if the improved scheme \eqnref{eqn:mubar scheme} is adopted.
Note that
\be
\label{eqn:rho crit infty}
3\rho_\textrm{Pl}=
\rho_\textrm{crit}^{(0)} < \rho_\textrm{crit}^{(1)} < \cdots < \rho_\textrm{crit}^{(\infty)}=\frac{3\pi^2}{4}\rho_\textrm{Pl}
\ee
This shows that the bouncing scenario is very similar to the $V(\phi)=0$ case as studied in \cite{Chiou:2009hk} and the condition for occurrence of the quantum bounce remains exactly the same.

Furthermore, in the immediate vicinity of the quantum bounce, we have $\cos(\mubar c)\approx 0$ and thus the second line of \eqnref{eqn:Raychaudhuri eq} dominates the first line, giving $\dot{H}>0$. Therefore, close to the quantum bounce, the quantum effects of holonomy corrections drive a short phase of super-inflationary expansion (i.e. $\dot{H}>0$). According to the analysis of \cite{Chiou:2009hk}, the quantum bounce takes place more abruptly for larger $n$ (see \figref{fig:zoom in}(d)). Thus we anticipate that the super-inflationary phase is shorter but more abrupt as $n$ increases (see \figref{fig:inflation}(c) and \figref{fig:zoom in}(c)).

\section{Inflation in the symmetric bouncing scenario}

The super-inflation can drive the inflation field up to its potential hill, thus initiating the standard inflation. This has been demonstrated in \cite{Tsujikawa:2003vr}, where loop quantum corrections on the eigenvalues of the geometrical density operator $\hat{d}_j$ are considered. Inflation in the framework of LQC with holonomy corrections has also been studied \cite{Mielczarek:2009zw}. In this paper, we investigate whether the standard inflation can be initiated by the quantum bounce resulting from (higher order) holonomy corrections. Since we have shown that there exists a super-inflationary phase close to the bounce, we anticipate the answer to be affirmative and expect a similar scenario to that in \cite{Tsujikawa:2003vr}.

Additionally, unlike \cite{Mielczarek:2009zw}, we cast the inflationary evolution in a \emph{symmetric} bouncing scenario, the merit of which is that the initial conditions can be naturally and uniquely set at the bouncing epoch once the (symmetry) potential is given (and thus no additional tuning parameters are needed except for those describing the potential).

To make this more precise, consider the simple potential:
\be\label{eqn:V}
V(\phi)=\frac{1}{2\hbar^2}m_\phi^2\phi^2.
\ee
In order to have a symmetry bouncing solution (i.e. pre-bounce and post-bounce universes are the mirrored solutions to each other), we must have
\be\label{eqn:phi0}
\phi(t=0)=:\phi_0=0,
\ee
where the epoch of the bounce is set at $t=0$. The condition of the quantum bounce \eqnref{eqn:rho crit} then uniquely set the initial velocity of $\phi$ as
\begin{subequations}\label{eqn:phi dot}
\ba
\label{eqn:phi dot 1}
\left.\dot{\phi}\right|_{t=0}
&=&\left[6\mathfrak{F}_n^2\rho_\mathrm{PL}-2V(\phi_0)\right]^{1/2}\\
\label{eqn:phi dot 2}
&\mathop{=}\limits_{\phi_0=0}&
\sqrt{6}\,\mathfrak{F}_n\rho_\mathrm{Pl}.
\ea
\end{subequations}
Additionally, we have $\mubar c=\pi/2$ at the bouncing epoch and thus
\be
c(t=0)
=\frac{\pi}{2}\sqrt{\frac{p(t=0)}{\Delta}}\,.
\ee
(Note that the heuristic effective dynamics is invariant under the rescaling $p\rightarrow l^2p$ together with $c\rightarrow l c$ if \eqnref{eqn:mubar scheme} is adopted. See \cite{Chiou:2007mg} for more details.)

Once $m_\phi$ is given, the initial conditions at the bouncing epoch is uniquely determined (up to an irrelevant rescaling factor $l$) and then the evolution govern by the Hamilton's equations \eqnref{eqn:Hamilton eqs} can be solved numerically (by the Runge-Kutta method). The initial conditions given above will yield a symmetric solution.

In this model, varying the value of $m_\phi$ gives different $\phi_{\max}$ (the maximal value of $\phi$ at the potential hill) and consequently yields different numbers of $e$-foldings. The values for different $m_\phi$ are listed in Table~\ref{tab:values}. It should be noted that unlike the standard slow-roll inflation, whereby both $m_\phi$ and $\phi_{\max}$ are tuning parameters, we now have $m_\phi$ be the \emph{only} tuning parameter, by which the value of $\phi_{\max}$ is automatically determined.

\begin{figure*}
\begin{picture}(500,300)(0,0)

\put(-65,-445)
{
\scalebox{1}{\includegraphics{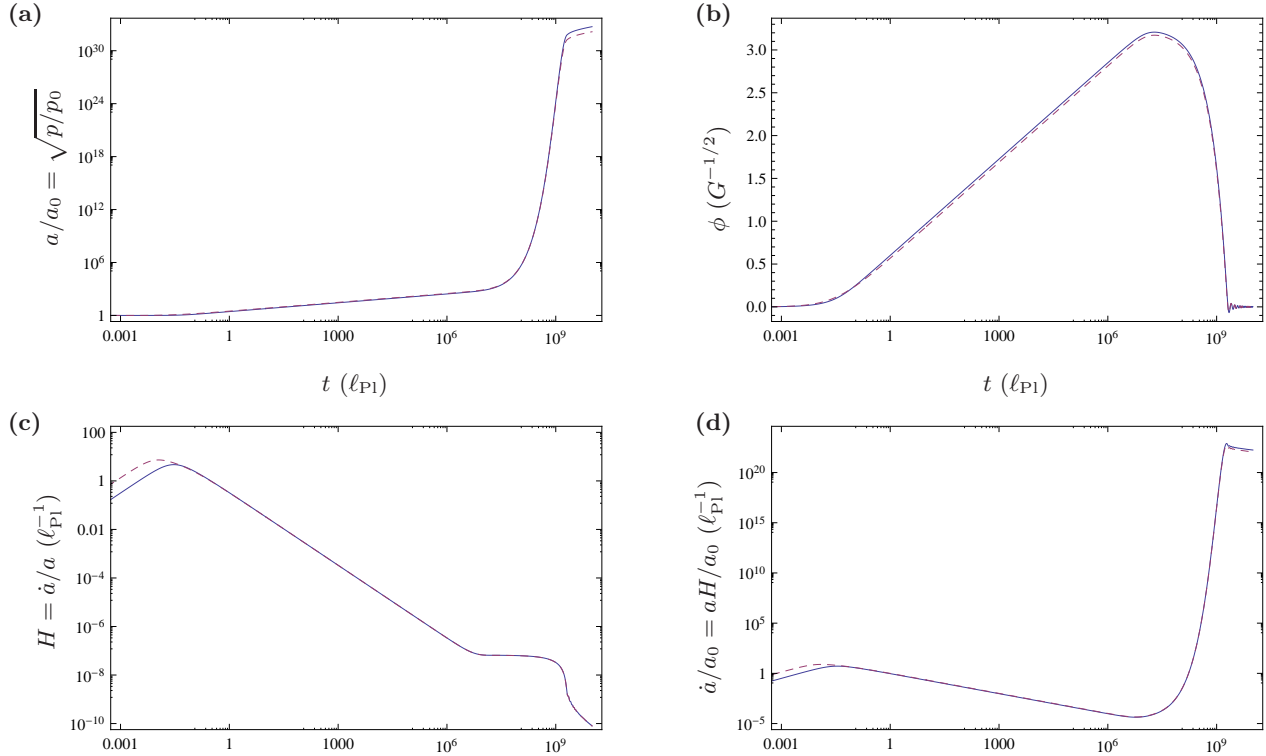}}
}

\end{picture}
\caption{Evolutions for the cases of $n=0$ (solid curve) and $n=100$ (dashed curve) with $m_\phi=10^{-8}M_\mathrm{Pl}$ and $\gamma=\ln2/(\sqrt{3}\,\pi)$. \textbf{(a)} Growth of $a=L^{-1}p$ with $a_0=L^{-1}p_0=a(t=0)$. The final $e$-foldings are about $\ln10^{32}$. \textbf{(b)} Evolution of $\phi$. After $\phi$ reaches $\phi_{\max}$, it starts to roll down slowly. In the end of the slow roll, $\phi$ oscillates. \textbf{(c)} Evolution of the Hubble rate $H$. We can see a short period of the super-inflationary phase ($\dot{H}>0$ or $H$ is increasing) close to the quantum bounce. \textbf{(d)} Evolution of $\dot{a}/a_0$. We can see a non-inflationary phase ($\ddot{a}<0$ or $\dot{a}$ is decreasing) between the super-inflation and the standard inflation.}\label{fig:inflation}
\end{figure*}

\begin{figure*}
\begin{picture}(500,300)(0,0)

\put(-65,-445)
{
\scalebox{1}{\includegraphics{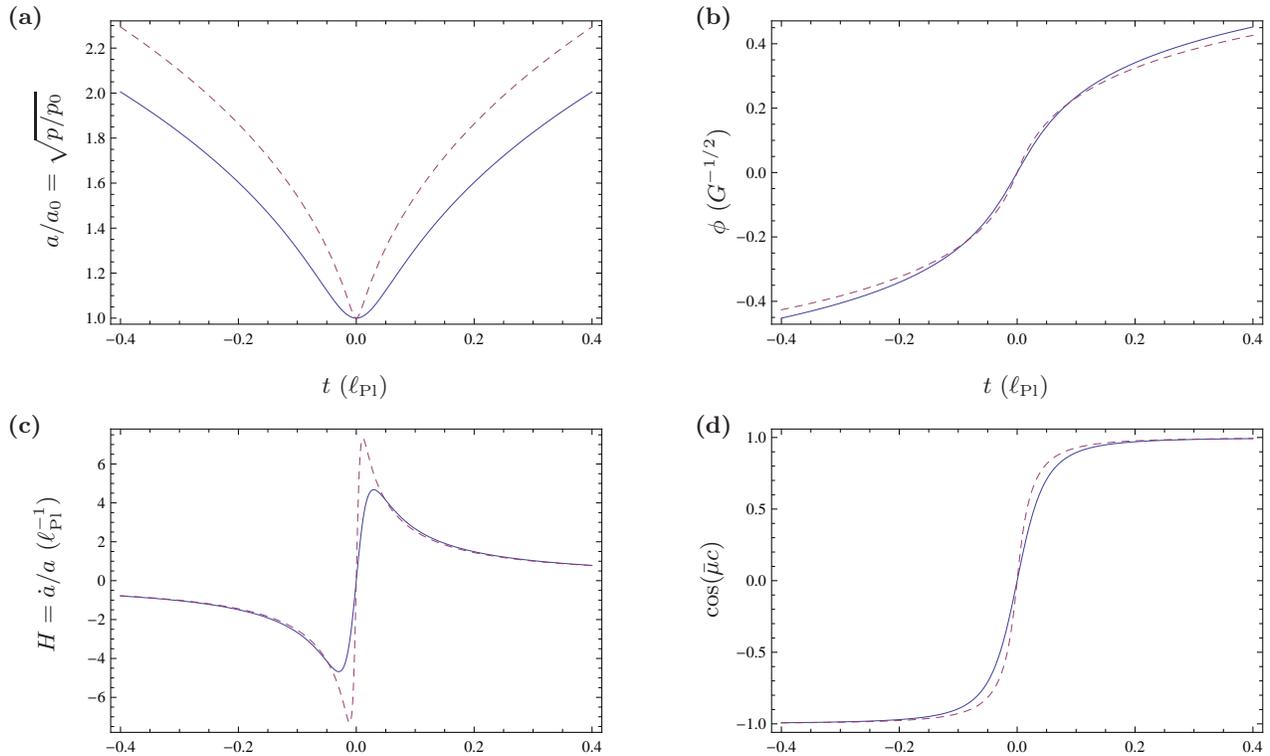}}
}

\end{picture}
\caption{Zoom in of \figref{fig:inflation} around the quantum bounce (with the pre-bounce side included). \textbf{(a)} Evolution of $a$. \textbf{(b)} Evolution of $\phi$. \textbf{(c)} Evolution of $H$. We can see that the super-inflationary phase ($\dot{H}>0$) is shorter and more abrupt for larger $n$. \textbf{(d)} Evolution of $\cos(\mubar c)$, signaling how significant the quantum effects are. $\cos(\mubar c)\rightarrow\pm1$ when the universe is classical.}\label{fig:zoom in}
\end{figure*}

To solve the horizon and flatness problems, around 70 $e$-foldings of inflation are required \cite{Liddle:2000cg}. Therefore, we choose $m_\phi=10^{-8}M_\mathrm{Pl}$ (the Planck mass $M_\mathrm{Pl}:=\sqrt{\hbar/G}$\,) for the detailed numerical analysis. The results are shown in \figref{fig:inflation}, and \figref{fig:zoom in} zooms in on the behaviors around the bouncing epoch (with the inclusion of the pre-bounce side).

\figref{fig:inflation} shows that there are two stages of inflation ($\ddot{a}>0$ or $\dot{a}=aH$ is increasing) with a non-inflationary phase in between. (Likewise, as symmetric to the post-bounce evolution, the pre-bounce universe undergoes the deflationary, non-deflationary and super-deflationary phases in succession and is then bridged to the post-bounce universe by the quantum bounce.)

The first stage of inflation is the short super-inflation ($\dot{H}>0 $ or $H$ is increasing) near the bouncing epoch as predicted in \eqnref{eqn:Raychaudhuri eq}. The super-inflation ends very rapidly since the quantum effects quickly become inconsiderable as can be seen in \figref{fig:zoom in}(d).

The second stage of inflation begins as $\phi$ reaches its maximum ($\dot{\phi}=0$) and starts to roll down slowly. Since quantum corrections are completely negligible ($\mubar c\ll 1$) at this stage, this is the standard slow-roll inflation, followed by conventional rehearing when the inflaton oscillates.

The two-stage inflationary scenario is in a very similar fashion to that studied in \cite{Tsujikawa:2003vr}. However, the super-inflation here is caused by the holonomy corrections of LQC, instead of the corrections on $\hat{d}_j$. Additionally, the evolution is now symmetric and the variant of it depends only on a single parameter $m_\phi$.

For a broad range of $m_\phi$, this scenario remains qualitatively unchanged but yields different $\phi_{\max}$ and $e$-foldings (see Table~\ref{tab:values}). For an appropriate value of $m_\phi$ around $10^{-7}$--$10^{-8}M_\mathrm{Pl}$, loop quantum effects of holonomy corrections can set adequate initial conditions for the standard inflation to achieve a desired number of $e$-foldings.

Also note that this scenario is robust for different $n$. As $n$ increases, the value of $\phi_{\max}$ and thus the number of $e$-foldings decrease only slightly as shown in \figref{fig:inflation}(a) and (b). For larger $n$, the super-inflationary phase is shorter and more abrupt as can be seen in \figref{fig:inflation}(c) and \figref{fig:zoom in}(c). In principle, the quantum bounce can be arbitrarily abrupt and brief as $n\rightarrow\infty$ at the level of heuristic effective dynamics, but in the quantum theory of LQC the abruptness will be well smoothed by the quantum fluctuations (see \cite{Chiou:2009hk,Chiou:2009yx} and Footnote~\ref{footnote:high order}).

Furthermore, as can be seen in \figref{fig:inflation}(d), between the super-inflation and the standard slow-roll inflation, there is a non-inflationary phase ($\ddot{a}<0$), which violates the slow-roll condition. The violation of slow roll is an indirect loop quantum gravity effect in the sense that it occurs after the super-inflation ends and the universe becomes classical. This is qualitatively the same as that in \cite{Tsujikawa:2003vr} and therefore we anticipate that violation of the slow-roll condition will lead to some suppression of the power spectrum at large scales and running of the spectral index in a fashion akin to that discussed in \cite{Tsujikawa:2003vr}. As $n$ increases, the super-inflation is shorter and thus the violation of slow-roll begins earlier. Therefore, different orders $n$ are expected to give different suppressions on the low angular power spectrum of CMB and in principle the very low angular powers can be strongly suppressed in the limit $n\rightarrow\infty$. The details are however more involved and we hope to investigate them in the future.\footnote{To investigate the quantum imprint on the CMB in the symmetric bouncing scenario, we have to first elaborate on the issue whether the quantum fluctuations should be set as symmetric about the bouncing epoch. In the quantum theory of LQC, it is permissable that the minima of the symmetric trajectories of $\langle p \rangle$ and $\langle p^2 \rangle-\langle p \rangle^2$ do not coincide (see \cite{Bojowald:2006gr} and \cite{Corichi:2007am}).}

\begin{table*}
\begin{tabular}{c|ccccc}
\hline
$m_\phi$ $(M_\mathrm{Pl})$ & $10^{-5}$ & $10^{-6}$ & $10^{-7}$ & $10^{-8}$ & $10^{-9}$ \\
\hline
$\phi_{\max}$ $(G^{-1/2})$ & $2.14$ & $2.50$ & $2.85$ & $3.21$ & $3.57$ \\
\hline
$e$-foldings & $\sim \ln10^{15}$ & $\sim \ln10^{20}$ & $\sim \ln10^{26}$ & $\sim \ln10^{32}$ & $\sim \ln10^{39}$\\
& $\approx35$ & $\approx46$ & $\approx60$ & $\approx74$ & $\approx90$\\
\hline
\end{tabular}
\caption{Values of $\phi_{\max}$ and $e$-foldings for different $m_\phi$ in the case of $n=0$.}\label{tab:values}
\end{table*}

\section{Summary and discussion}\label{sec:summary}
We have shown that, close to the quantum bounce caused by holonomy corrections of LQC, the universe undergoes a short super-inflationary phase, which can drive the inflaton field $\phi$ up to its potential hill $\phi_{\max}$, thus setting the initial conditions for the standard slow-roll inflation. The two-stage inflationary scenario with a non-inflationary phase in between is akin to that in \cite{Tsujikawa:2003vr}. Given the simple potential in \eqnref{eqn:V}, for a broad range of $m_\phi$, this scenario remains qualitatively the same but yields different values of $\phi_{\max}$ and $e$-foldings. With a proper value of $m_\phi$, the super-inflation can yield $\phi_{\max}\agt 3\,G^{-1/2}$ to achieve sufficient $e$-foldings. This scenario is robust for different orders $n$, which only slightly change the value of $\phi_{\max}$ and the number of $e$-foldings.

In the symmetric bouncing scenario, once the symmetric potential $V(\phi)$ is specified, the condition for occurrence of the quantum bounce naturally and uniquely fixes the initial conditions at the bouncing epoch and thus no extra tuning parameters are needed except for those describing $V(\phi)$. For the simple potential \eqnref{eqn:V}, $m_\phi$ is the only tuning parameter, in contrast to the standard slow-roll inflationary model, where both $m_\phi$ and $\phi_{\max}$ are tunable.

Instead of the natural choice \eqnref{eqn:phi0}, had we chosen $\phi_0\neq0$ at the bouncing epoch, we would have had a value different from \eqnref{eqn:phi dot 2} for $\dot{\phi}$ at $t=0$ and obtained an asymmetric inflationary scenario as discussed in \cite{Mielczarek:2009zw}. The asymmetric scenario is somewhat less natural in the sense that the pre-bounce universe can be completely different from the post-bounce universe and no first principle can be applied to specify the value of $\phi_0$. Other forms of the potential $V(\phi)$ can also be easily adopted in the symmetric bouncing scenario as long as $V(\phi)$ is symmetric, i.e. $V(-\phi)=V(\phi)$, although $V(\phi)$ may involve more tuning parameters. However, for some inflationary models (such as the power-law inflation, in which the potential is chosen to take the exponential form \cite{Liddle:2000cg}), $V(\phi)$ is asymmetric and cannot fits in the symmetric bouncing scenario.

We also note that the non-inflationary phase between the super-inflation and the standard inflation violates the slow-roll condition and thus is expected to lead to some suppression on the low angular power spectrum of CMB in the similar fashion as discussed in \cite{Tsujikawa:2003vr}. Different orders $n$ are expected to give different suppressions and in principle the very low angular powers can be strongly suppressed in the limit $n\rightarrow\infty$. Further investigation is needed for the details.


\begin{acknowledgements}
The authors would like to thank Abhay Ashtekar for suggesting this research topic. D.W.C. is grateful to Jiun-Huei Wu for the warm hospitality during his visit at National Taiwan University, where this paper was completed. This work was supported in part by the NSFC Grant No. 10675019 and the financial support Grant No. 20080440017 and Grant No. 200902062 from China Postdoctoral Science Foundation.
\end{acknowledgements}



\begin{thebibliography}{99}


\bibitem{Ashtekar:2006rx}
  A.~Ashtekar, T.~Pawlowski and P.~Singh,
  ``Quantum nature of the big bang,''
  Phys.\ Rev.\ Lett.\  {\bf 96}, 141301 (2006)
  [arXiv:gr-qc/0602086].

\bibitem{Ashtekar:2006uz}
  A.~Ashtekar, T.~Pawlowski and P.~Singh,
  ``Quantum nature of the big bang: An analytical and numerical investigation I,''
  Phys.\ Rev.\  D {\bf 73}, 124038 (2006)
  [arXiv:gr-qc/0604013].

\bibitem{Ashtekar:2006wn}
  A.~Ashtekar, T.~Pawlowski and P.~Singh,
  ``Quantum nature of the big bang: Improved dynamics,''
  Phys.\ Rev.\  D {\bf 74}, 084003 (2006)
  [arXiv:gr-qc/0607039].



\bibitem{Ashtekar:2006es}
  A.~Ashtekar, T.~Pawlowski, P.~Singh and K.~Vandersloot,
  ``Loop quantum cosmology of k = 1 FRW models,''
  Phys.\ Rev.\  D {\bf 75}, 024035 (2007)
  [arXiv:gr-qc/0612104].

\bibitem{Vandersloot:2006ws}
  K.~Vandersloot,
  ``Loop quantum cosmology and the k = -1 RW model,''
  Phys.\ Rev.\  D {\bf 75}, 023523 (2007)
  [arXiv:gr-qc/0612070].

\bibitem{Chiou:2006qq}
  D.~W.~Chiou,
  ``Loop quantum cosmology in Bianchi I models: Analytical investigation,''
  Phys.\ Rev.\  D {\bf 75}, 024029 (2007)
  [arXiv:gr-qc/0609029].

\bibitem{Chiou:2007mg}
  D.~W.~Chiou,
  ``Effective dynamics, big bounces and scaling symmetry in Bianchi type I loop quantum cosmology,''
  Phys.\ Rev.\  D {\bf 76}, 124037 (2007)
  [arXiv:0710.0416 [gr-qc]].

\bibitem{Chiou:2007sp}
  D.~W.~Chiou and K.~Vandersloot,
  ``The behavior of nonlinear anisotropies in bouncing Bianchi I models of loop quantum cosmology,''
  Phys.\ Rev.\  D {\bf 76}, 084015 (2007)
  [arXiv:0707.2548 [gr-qc]].



\bibitem{Bojowald:2002nz}
  M.~Bojowald,
  ``Inflation from Quantum Geometry,''
  Phys.\ Rev.\ Lett.\  {\bf 89}, 261301 (2002)
  [arXiv:gr-qc/0206054].

\bibitem{Bojowald:2003mc}
  M.~Bojowald and K.~Vandersloot,
  ``Loop quantum cosmology, boundary proposals, and inflation,''
  Phys.\ Rev.\  D {\bf 67}, 124023 (2003)
  [arXiv:gr-qc/0303072].


\bibitem{Date:2004yz}
  G.~Date and G.~M.~Hossain,
  ``Genericity of inflation in isotropic loop quantum cosmology,''
  Phys.\ Rev.\ Lett.\  {\bf 94}, 011301 (2005)
  [arXiv:gr-qc/0407069].

\bibitem{Tsujikawa:2003vr}
  S.~Tsujikawa, P.~Singh and R.~Maartens,
  ``Loop quantum gravity effects on inflation and the CMB,''
  Class.\ Quant.\ Grav.\  {\bf 21}, 5767 (2004)
  [arXiv:astro-ph/0311015].

\bibitem{Artymowski:2008sc}
  M.~Artymowski, Z.~Lalak and L.~Szulc,
  ``Loop Quantum Cosmology corrections to inflationary models,''
  JCAP {\bf 0901}, 004 (2009)
  [arXiv:0807.0160 [gr-qc]].

\bibitem{Mielczarek:2009zw}
  J.~Mielczarek,
  ``The Observational Implications of Loop Quantum Cosmology,''
  arXiv:0908.4329 [gr-qc].

\bibitem{Grain:2009kw}
  J.~Grain and A.~Barrau,
  ``Cosmological footprints of loop quantum gravity,''
  Phys.\ Rev.\ Lett.\  {\bf 102}, 081301 (2009)
  [arXiv:0902.0145 [gr-qc]].





\bibitem{Ashtekar:2007em}
  A.~Ashtekar, A.~Corichi and P.~Singh,
  ``On the robustness of key features of loop quantum cosmology,''
  Phys.\ Rev.\  D {\bf 77}, 024046 (2008)
  [arXiv:0710.3565 [gr-qc]].




\bibitem{Chiou:2009hk}
  D.~W.~Chiou and L.~F.~Li,
  ``How loopy is the quantum bounce? A heuristic analysis of higher order holonomy corrections in LQC,''
  Phys.\ Rev.\  D {\bf 79}, 063510 (2009)
  [arXiv:0901.1757 [gr-qc]].

\bibitem{Chiou:2009yx}
  D.~W.~Chiou and L.~F.~Li,
  ``Loop quantum cosmology with higher order holonomy corrections,''
  Phys.\ Rev.\  D {\bf 80}, 043512 (2009)
  [arXiv:0907.0640 [gr-qc]].




\bibitem{Bojowald:2006gr}
  M.~Bojowald,
  ``Large scale effective theory for cosmological bounces,''
  Phys.\ Rev.\  D {\bf 75}, 081301(R) (2007)
  [arXiv:gr-qc/0608100].

\bibitem{Chiou:2008bw}
  D.~W.~Chiou,
  ``Effective equations of motion for constrained quantum systems: A study of the Bianchi I loop quantum cosmology,''
  arXiv:0812.0921 [gr-qc].



\bibitem{Liddle:2000cg}
  A.~R.~Liddle and D.~H.~Lyth,
  \textit{\textit{Cosmological inflation and large-scale structure}},
  Cambridge University Press (2000).




\bibitem{Corichi:2007am}
  A.~Corichi and P.~Singh,
  ``Quantum bounce and cosmic recall,''
  Phys.\ Rev.\ Lett.\  {\bf 100}, 161302 (2008)
  [arXiv:0710.4543 [gr-qc]].

\end{thebibliography}
\end{document}